\documentclass[journal=jacsat,manuscript=article]{achemso}
\setkeys{acs}{articletitle = true}
\usepackage[version=3]{mhchem} 
\usepackage[T1]{fontenc}       

\author{M. Shcherbinin}
\affiliation[AU]{Department of Physics and Astronomy, Aarhus University, 8000 Aarhus C, Denmark}
\author{A. C. LaForge}
\affiliation[UF]{Physikalisches Institut, Universit{\"a}t Freiburg, 79104 Freiburg, Germany}
\author{M. Hanif}
\affiliation[AU]{Department of Physics and Astronomy, Aarhus University, 8000 Aarhus C, Denmark}
\author{R. Richter}
\affiliation[Unknown University]{Elettra Sincrotrone, 34149 Basovizza, Trieste, Italy}
\author{M. Mudrich}
\email{mudrich@phys.au.dk}
\affiliation[AU]{Department of Physics and Astronomy, Aarhus University, 8000 Aarhus C, Denmark}

\title[PIES]{Penning ionization of acene molecules by He nanodroplets}

\abbreviations{IR,NMR,UV}

\begin{document}

\begin{tocentry}

Some journals require a graphical entry for the Table of Contents.
This should be laid out ``print ready'' so that the sizing of the
text is correct.

Inside the \texttt{tocentry} environment, the font used is Helvetica
8\,pt, as required by \emph{Journal of the American Chemical
Society}.

The surrounding frame is 9\,cm by 3.5\,cm, which is the maximum
permitted for  \emph{Journal of the American Chemical Society}
graphical table of content entries. The box will not resize if the
content is too big: instead it will overflow the edge of the box.

This box and the associated title will always be printed on a
separate page at the end of the document.

\end{tocentry}

\begin{abstract}
Acene molecules (anthracene, tetracene, pentacene) and fullerene (C$_{60}$) are embedded in He nanodroplets (He$_N$) and probed by EUV synchrotron radiation. When resonantly exciting the He nanodroplets, the embedded molecules M are efficiently ionized by the Penning reaction $\mathrm{He}_N^*+\mathrm{M}\rightarrow\mathrm{He}_N + \mathrm{M}^+ + e^-$. However, the Penning electron spectra are broad and structureless -- showing no resemblance neither with those measured by binary Penning collisions, nor with those measured for dopants bound to the He droplet surface. The similarity of all four spectra indicates that electron spectra of embedded species are substantially altered by electron-He scattering. Simulations based on elastic binary electron-He collisions qualitatively reproduce the measured spectra, but require the assumption of unexpectedly large He droplets.
\end{abstract}

\section{Introduction}
He nanodroplets are widely used as cold and inert spectroscopic matrices of embedded `dopant' molecules and clusters~\cite{Toennies:2004,Stienkemeier:2006}. However, upon electronic excitation or ionization, He nanodroplets can induce severe perturbations of the spectra due to the interaction of the excited or ionized dopant with the surrounding He atoms~\cite{Buenermann:2007, Pentlehner:2010, TheisenJPCL:2011, Mudrich:2014}, or due to electron-He scattering~\cite{Wang:2008, Shcherbinin:2017}. To date, photoelectron spectroscopy has been employed by a few research groups for probing dopants in He nanodroplets and their relaxation dynamics~\cite{Radcliffe:2004, Loginov:2005, Loginov:2007, Loginov:2012, Loginov:2014, Fechner:2012}. In these studies, resonant multi-photon laser-ionization was applied mostly to metal atoms or clusters embedded in He nanodroplets. No photoelectron spectra of dopants by direct one-photon ionization have been reported so far. 

An alternative method to photoelectron spectroscopy is Penning ionization electron spectroscopy (PIES)~\cite{Siska:1993}. This method, which has been developed for many decades, has its merits for its sensitivity to the spatial electron distribution of molecules, clusters, and surfaces, and to anisotropic interaction potentials of the colliding reaction partners~\cite{Maruyama:2000}. Besides, Penning ionization experiments involving clusters can reveal additional details of complex ionization mechanisms such as autoionization of superexcited states~\cite{Tanaka:2000,Maruyama:2000} and diabatic relaxation of excitation~\cite{Wang:2008,Buchta:2013}. Traditionally, a rare gas atom, most often He due to its extremely high excitation energy, is prepared in an excited metastable state and collides with another atom, molecule, or surface, M, to induces the ionization of the latter in the reaction
\begin{equation}
\mathrm{He}^*+\mathrm{M}\rightarrow\mathrm{He} + \mathrm{M}^+ + e^-.
\end{equation}
In this reaction, the excess energy
\begin{equation}
E_e=E^*-E_i +\Delta E
\label{eq:Ee}
\end{equation}
is transferred to the emitted Penning electron $e^-$. Here, $E^*$ is the energy of the metastable rare gas atom, $E_i$ is the ionization energy of the colliding particle M, and $\Delta E$ is a small energy difference between potential energy curves of the incoming $\mathrm{He}^*+\mathrm{M}$ and outgoing $\mathrm{He}+\mathrm{M}^+ + e^-$ channels~\cite{Maruyama:2000}. Thus, by measuring the distribution of Penning electron kinetic energies, we obtain a spectrum of electron binding energies $E_i$ of M akin to its photoelectron spectrum (PES), provided $E^*$ is known and $\Delta E$ is known or negligible.

Penning ionization of molecules (SF$_6$) embedded in He nanodroplets was already reported in the pioneering study of the photoionization of large pure and doped He droplets using synchrotron radiation by the group of Toennies~\cite{Froechtenicht:1996}. When tuning the synchrotron to the most pronounced He droplet resonances around $h\nu=21.6$ and $23.8$~eV, increased yields of dopant ions detected. These observations were essentially reproduced by our earlier studies using alkali and earth-alkaline metal atoms as dopants~\cite{Buchta:2013, Mudrich:2014, LaForge:2016}. The higher yields measured for the latter dopants were rationalized mainly by the surface location of these dopants, which is favorable for Penning ionization given that the excited He$^*$ atom tends to be expelled out of the bulk toward the surface of the droplet due to repulsive He$^*$-He droplet interactions~\cite{Buchenau:1991, Kornilov:2011}. Note that Penning ionization of dopants is also seen in experiments using electron bombardment as a method of exciting doped He nanodroplets~\cite{Scheidemann:1997,Lan:2011,Lan:2012}. Besides, Penning ionization of molecules (benzene, benzonitrile, toluene, pyridine) attached to clusters made of the heavier rare-gases (neon, argon and krypton) has been reported~\cite{Kamke:1985,Kamke:1986,Kamke:1987}.

In the experiments using alkali metals and rare-gas atoms as dopants of He nanodroplets, also Penning ionization electron spectra (PIES) were measured~\cite{Buchta:2013,Wang:2008}. In the case of alkali metal dopants, the PIES are dominated by one well-defined peak near $E^*-E_i$, where $E^*=20.6~$eV (1s2s$^1$S-state of He) and $E_i$ is the ionization energy of the dopant atom. The PIES of Kr and Xe featured two pairs of peaks, indicating that Penning ionization of the rare-gas atom proceeded from He$^*$ either in the 1s2p$^1$P-state or in the 1s2s$^1$S-state, the latter being populated by droplet-induced relaxation. In addition, a broad feature reaching down to an electron energy $E_e=0$ was present, which dominated the spectrum when increasing the He droplet size to $N>10^4$. This feature was discussed in the context of electron-He scattering. However, we note that the atomic lines remained visible in the PIES at all experimental conditions.

The aim of this study is to present and discuss PIES of the molecular dopants anthracene (Ac), tetracene (Tc), pentacene (Pc), and fullerene (C$_{60}$) embedded inside He nanodroplets. Contrary to our earlier findings for surface-bound alkali metals and rare-gas atoms~\cite{Wang:2008, Buchta:2013}, the electron spectra are broad and nearly structureless, showing no resemblance with the respective gas phase PIES or PES. Thus, unfortunately, He nanodroplets appear not to be generally suitable for Penning electron spectroscopy, and we call those for caution who may have high expectations regarding the resolution of photoelectron or Penning electron spectra of molecules embedded inside He nanodroplets.

\section{Methods}
\label{sec:methods}
The setup used for the present experiments has been described previously~\cite{Buchta:2013,Shcherbinin:2017}. Briefly, a beam of He droplets with an average diameter of 6~nm is produced by continuously expanding pressurized He (50~bar) out of a cold nozzle (diameter 5~$\mu m$, temperature 14~K). The He droplets are doped with one molecule on average by pickup inside a heated vapor cell [length 1~cm, temperature 35$^\circ$ (Ac), 110$^\circ$ (Tc), 165$^\circ$ (Pc), 400$^\circ$ (C$_{60}$)]. At these conditions, the proportion of molecular dimers with respect to monomers in the mass spectra remain well below 10\,\%. Therefore we exclude substantial contributions of dopant oligomers to the detected electron and ion signals.

The EUV light beam at the Gasphase beamline of Elettra Sincrotrone, Trieste, is narrow-band ($\nu /\Delta\nu > 10^3$) and tunable over the discrete absorption bands of He nanodroplets up to the He ionization threshold~\cite{Joppien:1993}. In the photon energy range 19-23~eV we use a 0.2~$\mu m$ thick tin filter to suppress higher order radiation. Electrons and ions created by photoionization of the doped He nanodroplets are detected in coincidence using the photoelectron-photoion coincidence velocity-map imaging (PEPICO-VMI) technique~\cite{Buchta:2013,BuchtaJCP:2013}. Either electron or ion velocity-map images are recorded in correlation with ion masses. Inverse Abel transformation yields electron and ion kinetic energy spectra and angular distributions~\cite{Dick:2014}. To discriminate signals correlated with the He droplet beam from the background gas (mainly water and dopant molecules effusing out of the heated cell along the He droplet beam axis), we use a mechanical beam chopper which periodically blocks the He droplet beam.

To simulate the influence of a He droplet on an electron emitted from a dopant molecule located inside the droplet, we compute classical electron trajectories through the droplet, subjected to binary elastic collisions with He atoms. This approach is inspired by previous studies of the interaction of electrons with bulk liquid He, which showed that the relaxation of hot electrons was mainly governed by elastic binary collisions between the electron and individual He atoms~\cite{Onn:1971,Loginov:2005}. Electron-He scattering is implemented by a Monte-Carlo method based on doubly differential (energy, scattering angle) electron-He scattering cross sections~\cite{Adibzadeh:2005}. Assuming the initial Penning process to occur in the center of the He nanodroplet, the classical electron trajectory is calculated in three dimensions up to the droplet surface, while accounting for electron-He scattering and Coulomb interaction of the electron with the ion which remains fixed at the droplet center. The He number density inside the spherical droplet is taken as homogeneous with a value 0.022~\AA$^{-3}$~\cite{Harms:1998}. Assuming that the initial energy distribution of the Penning electrons is that of the PIES measured for gas phase Ac~\cite{Yamauchi:1998}, the simulation is repeated for $10^5$ electrons for each value of the initial electron energy and for different droplet radii. The final droplet PIES is given by the histogram of kinetic energies of those electrons that have escaped out of the He droplets.  

\begin{figure}
	\centering \includegraphics[width=0.75\textwidth]{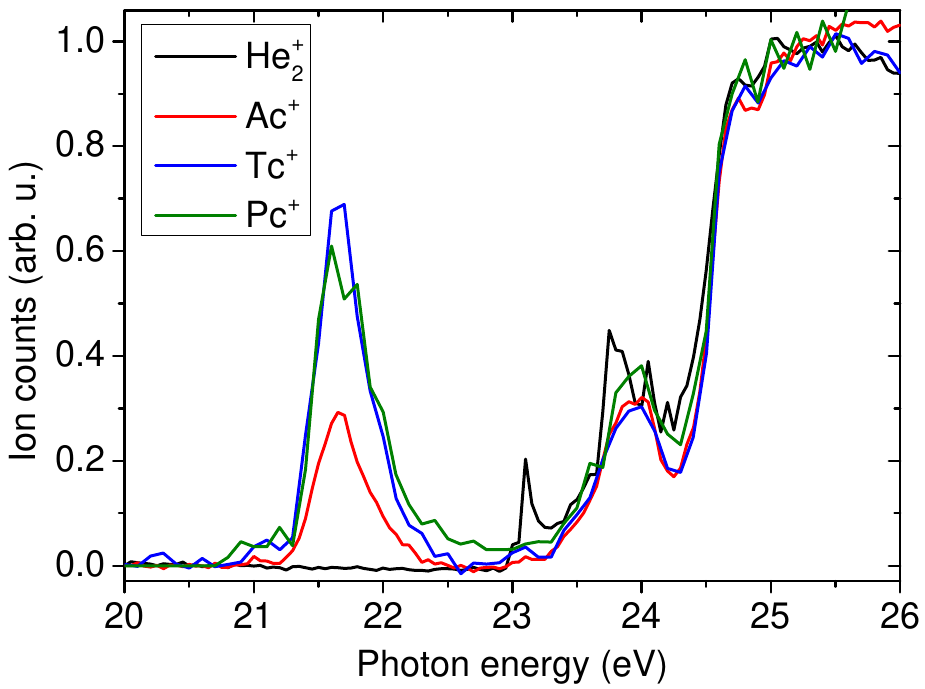}
	\protect\caption{\label{fig:hvscan}
		Ion yield spectra of He$_2^+$, anthracene (Ac), tetracene (Tc), and pentacene (Pc) ions in the photon energy range around the He droplet absorption bands up to the ionization threshold.}
\end{figure}
\section{Results}
Clear evidence for Penning ionization of dopant molecules is obtained by recording the yield of dopant ions while scanning the photon energy across the absorption resonances of He nanodroplets. Since fragmentation is nearly absent, we only present unfragmented dopant ion yields and the corresponding electron spectra. Fig.~\ref{fig:hvscan} shows the ion yield for Ac, Tc, Pc ions, as well as the yield of He$_2^+$ ions for reference. In the range of photon energies between $h\nu=23$~eV up to the ionization threshold of He atoms ($E_i=24.6$~eV), He$_2^+$ ions are created by autoionization of highly excited He droplets~\cite{Froechtenicht:1996,Peterka:2007}. For $h\nu>24.6~$eV, both free He atoms and He droplets are directly ionized and we detect mainly free He$^+$ and He$_2^+$ ions. 

In these two ranges of $h\nu$, the yields of dopant ions closely follow that of He$_2^+$. Whenever a He$^+$ or He$_2^+$ charge is created inside a He droplet, charge transfer to the dopant particles can compete with the formation of free He$_2^+$. This leads to the ejection of the bare dopant ion or of complexes of the dopant ion with a small number of He atoms attached to it~\cite{Buchta:2013}. Note, however, that the He$_2^+$ yield curve features sharper peak structures near atomic Rydberg states, which are not present in the dopant ion spectra. This indicates that He$_2^+$ has formed following He ionization either in the bulk of the droplets, or at the droplet surface. Here, He atoms are less perturbed and therefore the absorption spectrum more closely resembles that of free atoms. The dopant ions, however, follow the broadened absorption profile of the droplet bulk~\cite{Joppien:1993}, consistent with their location in the interior of the droplets.

Moreover, the ion yields of all three acenes feature a clear maximum around $h\nu=21.6~$eV, which corresponds to the strongest droplet absorption resonance and correlates to the 1s2p$^1$P-state of the He atom. The yield of C$_{60}^+$ ions (not shown) at $h\nu=21.6~$eV is slightly lower than for the acenes, and amounts to 20\,\% of that at $h\nu=25~$eV. However, in contrast to Penning ionization of alkali metal ions~\cite{Buchta:2013}, the acene and C$_{60}^+$ ion yields stay well below those measured in the range where charge transfer ionization is active. The dopant ion signals at $h\nu=21.6~$eV in proportion to those at $h\nu>24.6~$eV are similar to those of alkaline earth dopants, which are located more deeply inside the He droplets~\cite{Mudrich:2014,LaForge:2016}. However, compared to other molecular dopants which have been studied so far (methane, fluorinated derivatives thereof, methanol, SF$_6$~\cite{Froechtenicht:1996,Peterka:2006}, Ne, Ar, Kr, Xe clusters~\cite{Kim:2006, Wang:2008, Buchta:2013}), the yields of Ac, Tc, and Pc Penning ions are the highest observed so far. 

\begin{figure}
	\centering \includegraphics[width=0.75\textwidth]{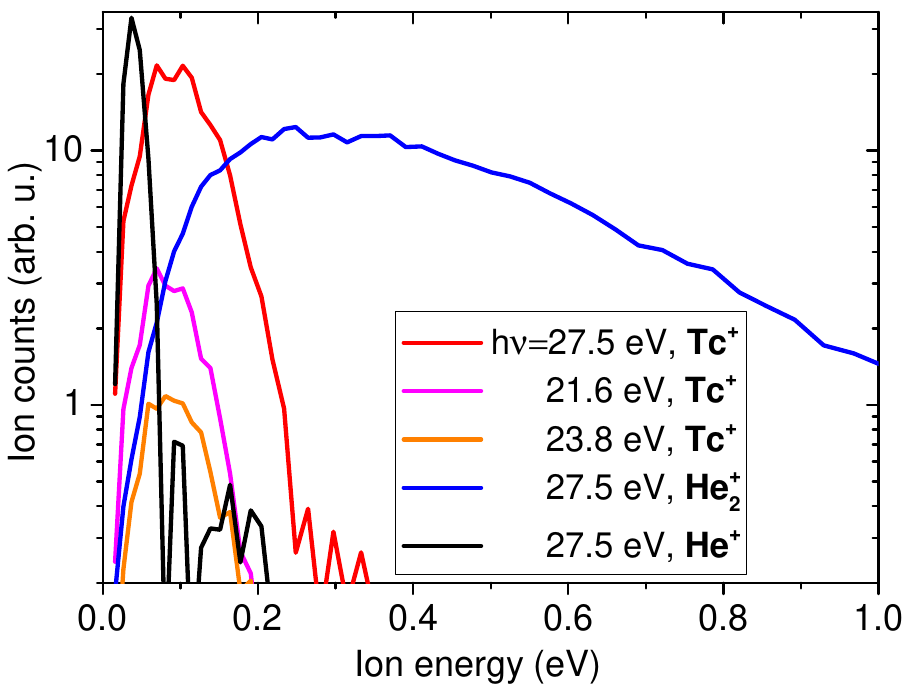}
	\protect\caption{\label{fig:ions}
		Kinetic energy distributions of He and Tc dopant ions generated by Penning ionization ($h\nu=21.6~$eV) by charge transfer ionization ($h\nu=26~$eV).}
\end{figure}
To obtain more detailed insight into the Penning reaction occurring inside the He nanodroplets, we have measured kinetic energies of the dopant ions. Fig.~\ref{fig:ions} shows the ion kinetic energy distributions of Tc ions recorded at various photon energies by operating the VMI spectrometer in ion-imaging mode. For reference, we include the kinetic energy spectra of He$^+$ and He$_2^+$ measured above the He ionization threshold ($h\nu=27.5~$eV). The angular distributions for all ions are fully isotropic. 

The kinetic energy of He$^+$ falls below 0.05~eV, which corresponds to the detection limit of our spectrometer at the used voltage setting. This very low kinetic energy is in line with our previous conclusion that free He$^+$ atomic ions cannot be emitted from singly ionized He droplets for energetic reasons. Instead, the measured He$^+$ ions originate from free He atoms that accompany the droplet beam~\cite{BuchtaJCP:2013}. In contrast, the He$_2^+$ ions are ejected out of the droplets with a finite energy around $0.3~$eV, driven by vibrational relaxation of He$_2^+$ into the ground state~\cite{BuchtaJCP:2013,Shcherbinin:2017}. For the Tc$^+$ ions, we measure kinetic energies around 0.1~eV, where charge transfer ionization ($h\nu=27.5~$eV) generates a slightly higher energy compared to Penning ionization ($h\nu<24~$eV). 

\begin{figure}
	\centering \includegraphics[width=0.7\textwidth]{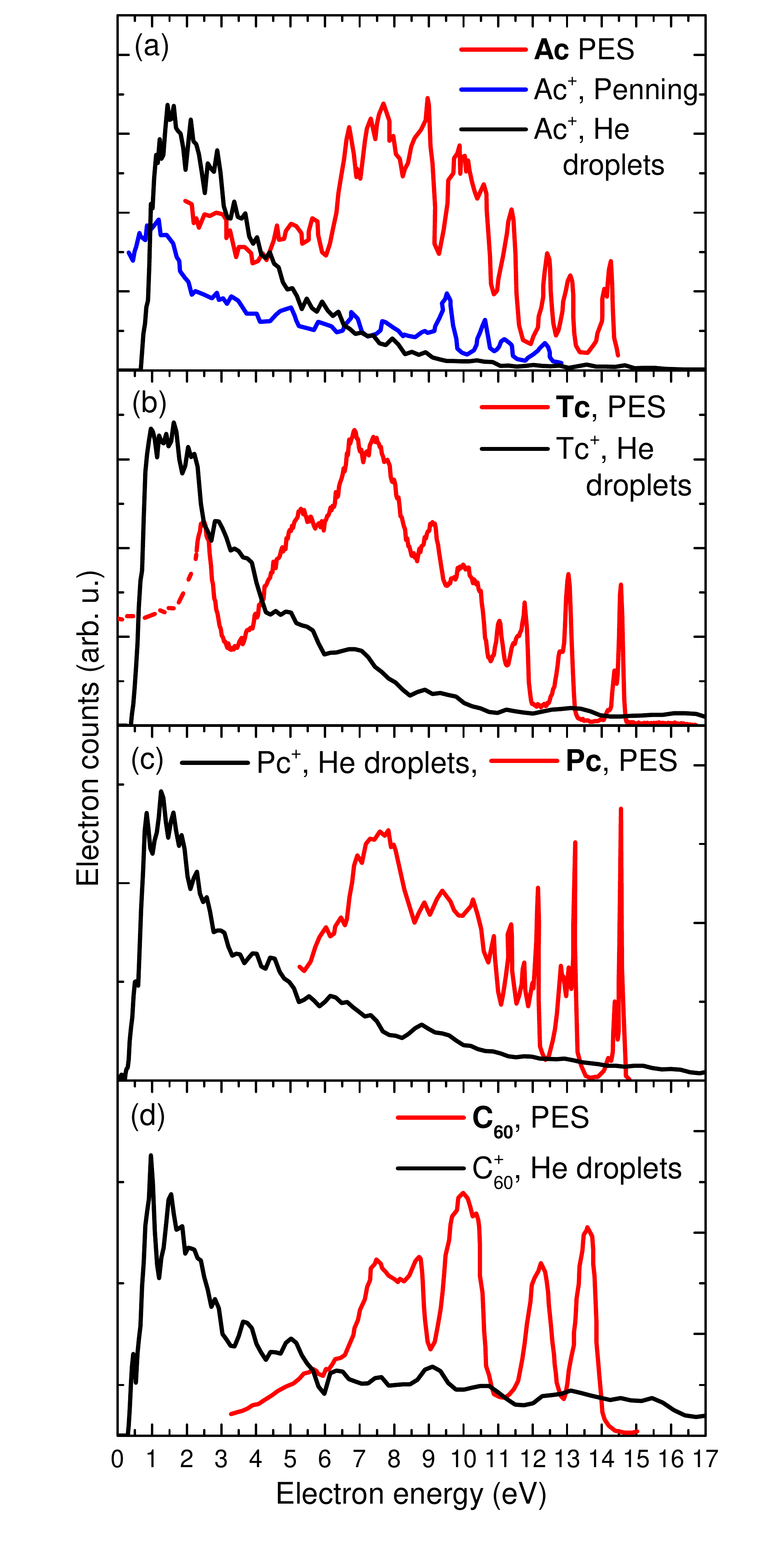}
	\protect\caption{\label{fig:PES}
		Comparison of gas phase photoelectron spectra (red lines)~\cite{Boschi:1972,Lichtenberger:1991,Yamauchi:1998} and Penning electron spectra (black lines) for Ac (a), Tc (b), Pc (c), and C$_{60}$ (d) embedded in He nanodroplets. Panel (a) includes the gas phase Penning electron spectrum of Tc~\cite{Yamauchi:1998}.}
\end{figure}
Given the relatively high yields of Penning ions detected for the acene dopants in He nanodroplets, we are in a position to record the corresponding PIES using the PEPICO-VMI technique. Fig.~\ref{fig:PES} displays a compilation of He droplet PIES measured at a fixed photon energy $h\nu=21.6~$eV for Ac (a), Tc (b), Pc (c), and C$_{60}$ (d) together with gas phase PES recorded with He-I line radiation. The latter PES are extracted from Refs.~\cite{Boschi:1972,Lichtenberger:1991,Yamauchi:1998}. Only the PES of Tc is measured in this work using a hemispherical electron analyzer and a dilute effusive beam of Tc. For this, the photon energy is set to $h\nu=20~$eV (solid red line). The low-energy part of the spectrum (dashed red line) is taken from the PES recorded at $h\nu=30$~eV. For Ac (a), the gas phase PIES was previously measured using crossed atomic beams (blue line)~\cite{Yamauchi:1998}. It strongly resembles the PES when taking into account the energetic down-shift due to the difference (1.4~eV) between $h\nu=21.2~$~eV of the He-I line and the energy of the metastable He($^3$S) atom inducing Penning ionization, $19.8~$eV. 

The PES feature complex peak structures which have been interpreted using electronic structure calculations~\cite{Boschi:1972,Lichtenberger:1991,Yamauchi:1998}. In contrast, the He droplet PIES are broadened toward low energies and nearly structureless. Moreover, the droplet PIES for the four species are very similar to one another up to different levels of signal-to-noise ratio. Note that these spectra are nearly independent of the He droplet size, the level of doping, and the photon energy in the range $h\nu=21$-$24~$eV. The most notable features are a signal maximum around 1.5~eV and a vanishing signal at $E_e<0.5$~eV. The angular distribution of droplet Penning electrons is isotropic [$\beta=0.0(2)$], in agreement with previous measurements~\cite{Wang:2008, Buchta:2013}.

\section{Discussion}
The high efficiency of Penning ionization of the acenes compared to most other dopants is likely related to their larger sizes, offering more contact points for He$^*$ to approach the dopant molecule before being ejected towards the droplet surface. Accordingly, the Penning signal of Tc and Pc (4 and 5 benzene rings, respectively) is higher than that of Ac (3 benzene rings). 
Additionally, the delocalized conjugated electron system of these aromatic molecules, which accounts for their large absorption cross sections in the visible spectral region, may also facilitate Penning ionization. 

Aside from the efficiency of ionization, the efficiency of ejection of the ions out of the droplets is an equally important factor determining the yield of detected free ions. Thus, the high yields of free ions may also be related to the degree of internal excitation of Penning ions, which facilitates ion ejection~\cite{SmolarekEjection:2010}. Note that the measured kinetic energy distributions (Fig.~\ref{fig:ions}) significantly deviate from the Maxwell-Boltzmann distribution (not shown). The latter features a sharp rise starting from zero energy and an extended falling edge toward high energies, whereas the measured distributions are peaked at a finite energy value (0.1~eV). This finding is in contrast to results obtained for molecular ions ejected from He nanodroplets by infrared excitation. In that case, the ion velocity distributions perfectly matched the Maxwell-Boltzmann distribution~\cite{SmolarekEjection:2010,ZhangJCP:2012}. Thus, in our case of indirect ionization by the He droplets, apparently a more impulsive ejection occurs compared to the ejection following laser-excitation of thermalized ions. While it is generally accepted that ions are ejected from He nanodroplets by non-thermal energy dissipation, a detailed understanding is still lacking. Further systematic studies of dopant species of different sizes, mass, atomic and electronic structures, and locations with respect to the He droplet surface will help elucidating this point.

The salient result of the present study is the extremely broadened and shifted PIES of Ac, Tc, Pc and C$_{60}$ in He droplet, which show no resemblance with the gas phase spectra. This must be related to a massive perturbation of any of the quantities on the right-hand side of Eq.~\ref{eq:Ee}, or by a modification of electron energies after the Penning reaction. Indeed, there are indications that $E^*$ undergoes ultrafast relaxation from the initial 1s2p$^1$P-state (21.6~eV) down to the atomic level 1s2s$^1$S (20.6~eV)~\cite{Wang:2008,Buchta:2013}. Assuming that Penning ionization occurs at all intermediate stages of the He$^*$ relaxation, this may account for a down-shifting of electron energies by up to 1~eV. This is insufficient for explaining our measurements, though. While $\Delta E$ is not expected to be notably affected by the He droplet compared to the gas phase, $E_i$ is known to be shifted inside He droplets due to the polarization effect of He surrounding the nascent Penning ion. For aniline molecules embedded in He nanodroplets, this shift was found experimentally and theoretically to be of the order of 0.1~eV, causing a slight up-shift of detected electron energies~\cite{Loginov:2005}. However, it is highly unlikely that much larger shifts with opposite sign should occur for the molecules studied here. 


\begin{figure}
	\centering \includegraphics[width=0.75\textwidth]{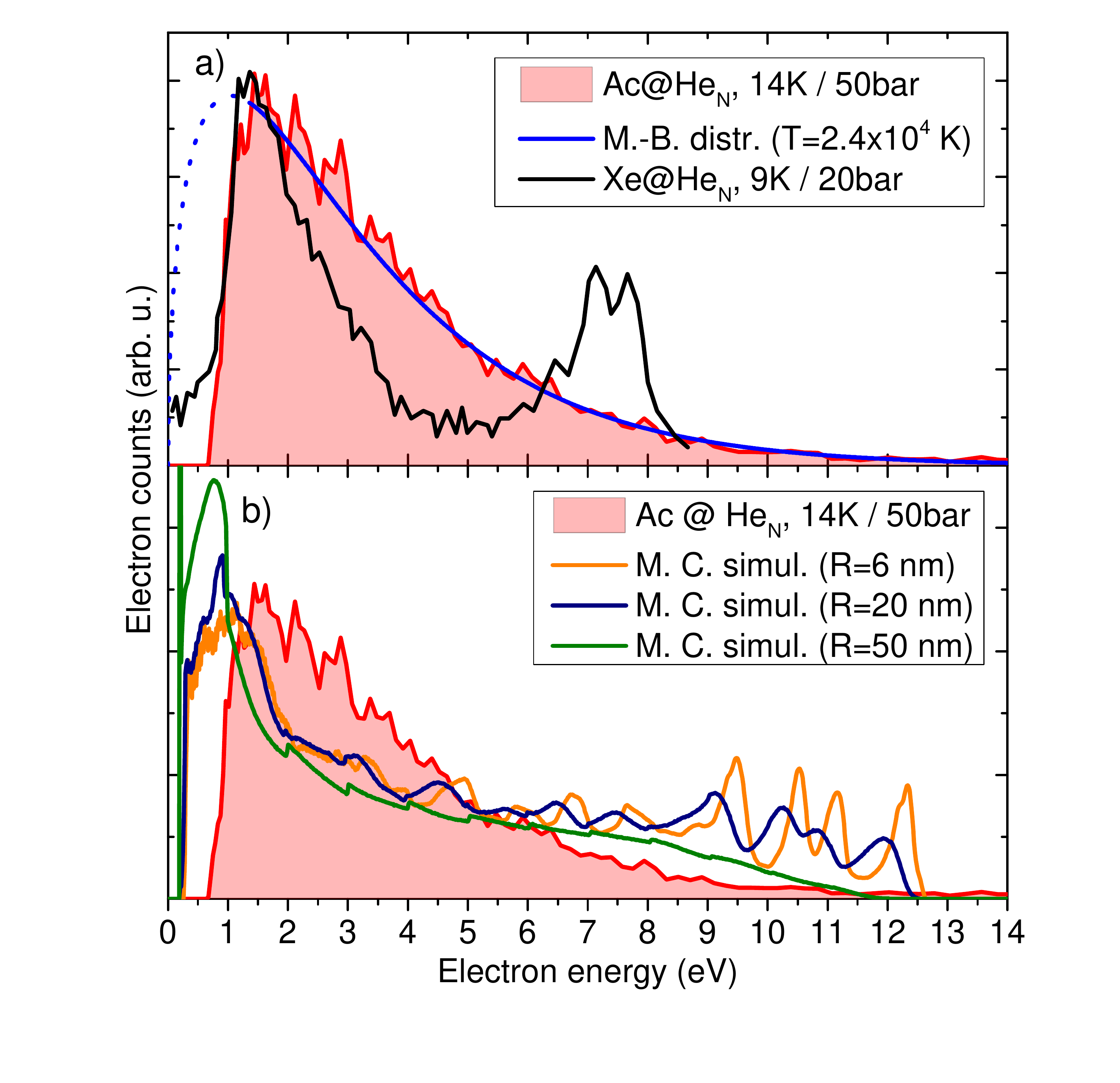}
	\protect\caption{\label{fig:Compare}
		Comparison between the experimental Tc (red line, this work) with a) Xe Penning electron spectra (black line, taken from Ref.~\cite{Wang:2008}) and a Maxwell-Boltzmann distribution (blue line). b) Experimental Tc spectrum and simulated spectra assuming electron-He scattering and various He droplet radii $R$.}
\end{figure}
Thus, the interaction of the Penning electron with the He droplet after the Penning reaction seems to be the dominant effect leading to the massive loss of electron energy~\cite{Wang:2008}. The fact that all PIES are quite similar in spite of the varying structure of the gas phase PIES and PES supports this conjecture. Electron-He scattering was previously found to perturb PES of embedded molecules~\cite{Loginov:2005,Wang:2008}. However, peak broadenings were only in the range $<0.15$~eV. Only PIES of Kr and Xe doped into the interior of He nanodroplets displayed a similarly broad feature as the one we see in the present work, as shown in Fig.~\ref{fig:Compare} a) (black line, extracted from Ref.~\cite{Wang:2008}). However, for all experimental conditions, the peak structure indicative for atomic-like Penning ionization prevailed, contrary to the present finding.

Furthermore, a sharp drop of electron signal at energies $<1~$eV was observed under conditions, where very large He droplets are formed by super-critical expansion, see the black line in Fig.~\ref{fig:Compare} a). The signal cut-off is likely due to the localization of the electron followed by electron-ion recombination~\cite{Loginov:2005,Wang:2008}. The cut-off energy of 1~eV matches the conduction band edge for electrons in liquid He~\cite{Asaf:1986,Buchenau:1991,Wang:2008}. For large He droplets, this band edge represents a barrier for the promotion of an electron into the conduction band, that is a state where it can freely move through the droplet and escape from it~\cite{Rosenblit:1995,Wang:2008}. Surprisingly, all our PIES feature a similar cut-off at energies $<0.5$-$1~$eV, although the average size of He nanodroplets generated at our experimental conditions (radius $R=6~$nm~\cite{Toennies:2004,Stienkemeier:2006}) would be considered as insufficient for the conduction band to be fully developed. In contrast, in the Xe experiments, a droplet size of $R=28~$nm was used~\cite{Wang:2008}. 

To assess our hypothesis that the observed PIES are mainly determined by electron-He scattering, we carry out Monte-Carlo simulations as outlined in the ``Methods'' section. Fig.~\ref{fig:Compare} b) includes the result of the simulation for three selected He droplet sizes as green lines. At the average droplet radius $R=6$~nm the peak structure of the initial $E_e$ distribution (PIES of Ref.~\cite{Yamauchi:1998}) is hardly altered. Only when $R$ increases to around $50$~nm, the simulated spectrum resembles the experimental one in the range $E_e>1.5~$eV. The large deviation at $E_e<1.5~$eV is due to the trapping of the electron below the conduction band edge and the subsequent electron-ion recombination as discussed above. In the simulation, electron-ion recombination occurs only at $E_e<0.2~$eV because neither many-body effects, nor quantum effects such as Pauli repulsion acting between the electron and the He atoms are taken into account. 

The fact that large He droplets ($R>10$~nm) are needed to achieve a similar degree a broadening in the simulation as in the experiment, as well as the occurrence of a cut-off energy in the experimental PIES, seem to indicate that the He droplets in our experiments are much larger than expected. Alternatively, the same results would be expected if our detection scheme were much more sensitive to the large He droplets component of the broad distribution of droplet sizes. However, from the characteristics of our droplet apparatus as a function of He nozzle temperature (pressures in the vacuum chambers, electron and ion signals), we infer with great certainty that droplet formation occurs in the sub-critical regime, where $R<10~$nm. Likewise, we have no reason to assume that large droplets contribute disproportionately to the measured yield of electrons and ions. On the contrary, electron-ion recombination in large droplets should reduce the detection efficiency. Thus, we argue that in the relevant energy range of 0-10~eV, the actual electron-He interaction is drastically underestimated by a model based solely on binary collisions. In comparison with Kr and Xe dopants, for which a contribution of weakly perturbed Penning electrons remained at all conditions, presumably the molecules studied here are more strongly localized at the droplet center due to a higher coordination number with the surrounding He so that an electron emitted from the molecule always has to pass through a layer of He before leaving the droplet. Clearly, more experimental and theoretical studies are needed to gain a better understanding of ionization processes in He droplets and the subsequent dynamics of emitted electrons and ions.

Finally, we mention that the falling edge of the PIES can be very well modelled by a Maxwell-Boltzmann distribution, as shown for Ac dopants by the blue line in Fig.~\ref{fig:Compare} a). This corresponds to a thermal electron distribution at a temperature of 24,000~K (2.1~eV), truncated at $E_e<1~$eV due to electron-ion recombination. While electron-He scattering should eventually lead to thermalization of the electron, clearly the temperature is incompatible with that of the He droplet (0.4~K). Extreme local heating of the environment around the electron would have to be invoked, which appears highly unlikely, though. 

\section{Conclusion}
In conclusion, we reported efficient Penning ionization of the acene molecules Ac, Tc, Pc, as well as C$_{60}$ doped into He nanodroplets. The Penning ion kinetic energy distribution of Tc$^+$ is peaked around 0.1~eV indicating impulsive ejection, contrary to previous measurements for laser-induced ejection of thermalized ions. Penning electron spectra are massively broadened toward low energies and feature a pronounced signal cut-off at electron energies below 1~eV, in contrast to the corresponding gas phase spectra and to previously measured He droplet Penning electron spectra of surface-bound dopants. Simulations based on electron-He binary scattering only reproduce the experimental results when assuming unexpectedly large He droplets, which indicates that the electron-He interaction in the relevant energy range is much more effective.

These results show that electron spectroscopy of dopants embedded inside He nanodroplets is not generally applicable. Further systematic studies for different types of dopants and different conditions for generating and doping the He droplet beam are required to fully characterize the ionization dynamics of dopants inside He droplets. In particular, the interaction of an electron with the He droplet deserves further investigation at energies where He polarization effects set in and eventually the electron localizes and recombines with the ion.

\begin{acknowledgement}

The authors thank Thomas Fennel for providing them with the code for the electron scattering simulation. Financial support by Aarhus Universitets Forskningsfond (AUFF) and by Deutsche Forschungsgemeinschaft (DFG, MU 2347/10-1) is gratefully acknowledged. A. C. L. gratefully acknowledges support by the Carl-Zeiss-Stiftung.

\end{acknowledgement}


\providecommand{\latin}[1]{#1}
\makeatletter
\providecommand{\doi}
{\begingroup\let\do\@makeother\dospecials
	\catcode`\{=1 \catcode`\}=2 \doi@aux}
\providecommand{\doi@aux}[1]{\endgroup\texttt{#1}}
\makeatother
\providecommand*\mcitethebibliography{\thebibliography}
\csname @ifundefined\endcsname{endmcitethebibliography}
{\let\endmcitethebibliography\endthebibliography}{}

\end{document}